\documentclass[manuscript]{aastex}

\slugcomment{CCS and HC$_3$N in L1147}

\shorttitle{CCS and HC$_3$N in L1147}
\shortauthors{Suzuki, Ohishi and Hirota}

\begin{document}


\title{Distribution of CCS and HC$_3$N in L1147, an Early Phase Dark Cloud}


\author{Taiki SUZUKI\altaffilmark{1}, Masatoshi OHISHI\altaffilmark{1, 2} and Tomoya HIROTA\altaffilmark{1, 2}}
\affil{$^1$ Department of Astronomical Science, the Graduate University for Advanced Studies (SOKENDAI), Osawa 2-21-1, Mitaka, Tokyo 181-8588, Japan}
\affil{$^2$ National Astronomical Observatory of Japan, Osawa 2-21-1, Mitaka, Tokyo 181-8588, Japan}

\email{taiki.suzuki@nao.ac.jp}




\begin{abstract}
We used the Nobeyama 45~m radio telescope to reveal spatial distributions of CCS and HC$_3$N in L1147, one of carbon-chain producing regions (CCPRs) candidates, where carbon-chain molecules are dominant rather than NH$_3$.
We found that three cores (two CCS cores and one HC$_3$N core) exist along the NE-SW filament traced by the 850~$\mu$m dust continuum, which are away from a  Very Low Luminosity Object (VeLLO -- a source that may turn into sub-stellar mass brown dwarf).
%
%
The column densities of CCS are 3--7$\times10^{12}$~cm$^{-2}$ and those of HC$_3$N are 2--6$\times10^{12}$~cm$^{-2}$, respectively, much lower than those previously reported towards other CCPRs.
%
%
%
%
We also found that two CCS peaks are displaced from that of HC$_3$N.
%
%
%
In order to interpret such interleaved distributions, we conducted chemical reaction network simulations, and found that
slightly different gas densities could lead to large variation of CCS-to-HC$_3$N ratio in the early phase of dark cloud evolution. 
Such a chemical ``variation'' may be seen in other CCPRs.
Finally we were able to confirm that the L1147 filament can be regarded as a CCPR.
%
%
%
%
\end{abstract}


\keywords{ISM: individual (L1147) --- ISM: molecules --- ISM: radio lines}



\section{Introduction}
It is well known that dark cloud cores, where low mass stars are formed, can be traced by abundant carbon-chain molecules.
\cite{Suzuki92} conducted survey observations of CCS, HC$_3$N, HC$_5$N, and NH$_3$ toward 49 molecular clouds, and pointed out that CCS is observed in the early phase of cloud evolution, while abundance of NH$_3$ gradually increases in the later phase.
The column density ratio of NH$_3$ to CCS (hereafter we call it as ``NH$_3$/CCS'' ratio) is a good indicator of cloud age, and clouds with low ``NH$_3$/CCS'' ratio are called Carbon-Chain Producing Regions (CCPRs).
Such characteristics indicate that the chemical and the physical evolution of a molecular cloud are well connected, providing with useful information in investigating the cloud evolution.
Since then, CCS observations were carried out to study the early phase cloud evolution. \citep[e.g.,][]{Hirahara92, Velusamy95, Ohashi99, Hirota02, Hirota04, Hirota06, Hirota09, Hirota11, Tafalla04}.

\cite{Hirota09} carried out survey observations in order to find CCPRs outside of the Taurus region, and suggested that five molecular clouds showed lower ``NH$_3$/CCS'' ratio less than 10, which were consistent with the known features of CCPRs.
Especially, four clouds, L1517D, L530D, L1147, and L1172B, showed a notable feature that their CCS column densities are lower than other CCPRs.
%

%
%
L1147 is a dark cloud in the Cepheus region at a distance of 325~pc \citep{Straizys92}. Since the original Lynds's catalog gave the same center positions for both L1147 and L1148 \citep{Lyn62}, some literatures call this dark cloud as L1148 (e.g., \cite{Kirk09}). Such a situation is confusing, however, we call the source as L1147 throughout this paper since we have already adopted this source name in our past papers.
\cite{Kirk05} obtained a dust continuum map towards L1147 taken by the SCUBA at 850~$\micron$, and reported that the dust continuum emission is weak in this cloud, which may represent physically young evolutionary phase.
\cite{Kirk09} also reported the dust continuum map of the entire region including L1147.
They mentioned that L1147 has a Very Low Luminosity Object (VeLLO),
which is defined as a source whose luminosity is lower than 0.1~$L_{\odot}$,
 and that it may form sub-stellar mass brown dwarfs \citep{Kauffmann11, Francesco06}.
This VeLLO source was discovered by the Spitzer mission, as Spitzer Multiband Imaging Photometer for Spitzer (MIPS) source, SSTc2d J204056.66+672304.9.

Among the CCPR candidates L1147 shows the lowest ``NH$_3$/CCS'' ratio of 1.5, however, \cite{Hirota09} observed a position away from the VeLLO source.
Thus further observation was needed to confirm that this dark cloud is certainly chemically young, since \cite{Hirota09} made single point observations for the CCPR candidate sources and did not exclude a possibility that these candidate sources are located at the edge of a dense core.

In this paper, we report mapping results towards L1147 in the 45~GHz CCS and HC$_3$N lines, and discuss its evolutionary phase using a chemical reaction network. The observational method is described in chapter 2 (Observations); intensity distributions of CCS and HC$_3$N, velocity structure as well as their column density distributions are shown in chapter 3 (Results); in chapter 4 (Discussion) we will compare the CCS and HC$_3$N abundances with our unpublished NH$_3$ data and chemical reaction simulation results to discuss if L1147 is chemically young CCPR; and finally we summarize our study results in chapter 5.

\section{Observations}


We conducted mapping observations with the 45~m radio telescope of the Nobeyama
Radio Observatory, National Astronomical Observatory of Japan, in April 2012. 
The CCS ($J_N$=$4_3$-$3_2$) and the HC$_3$N ($J$=5-4) lines in the 45~GHz region were observed simultaneously 
with a superconductor-insulator-superconductor (SIS) mixer receiver, whose system temperature was 200--340~K 
during our observations.
The main beam efficiency ($\eta_{\rm mb}$) was 0.72, and the beam size (FWHM) was 38".
The pointing accuracy was checked every about two hours by observing the SiO masers ($v$=1 and 2, $J$=1-0) towards T~Cep. 
The pointing error was estimated to be within 5''.
The rest frequencies, the intrinsic line strengths and the permanent dipole moments for CCS and HC$_3$N are $\nu$ = 45379.033 and 45490.316~MHz,
$S_{\rm ul}$=3.97 and 5.00, $\mu$ = 2.81 and 3.72 Debyes \citep{Murakami90,Lafferty78}, respectively.
Acousto-optical radio spectrometers with the frequency resolution of 37~kHz, 
corresponding to the velocity resolution of 0.24~km~s$^{-1}$, were used for the backend.
%
The mapping center was set to $\alpha_{2000}$=20$^{\rm h}$40$^{\rm m}$32$^{\rm s}$.0, $\delta_{2000}$=67$\degr$21'45''.
In the mapping observations of HC$_3$N and CCS, the spectra ware observed with a grid spacing of 40" in the position-switching mode, in which an offset position was 10' away from each source position.
We observed at 39 positions along the L1147 filament (for the mapping area, see, e.g., Figure~3).
The integration time per position was 200 seconds.



\section{Results}
\subsection{Molecular Distributions}

\begin{figure}
\plotone{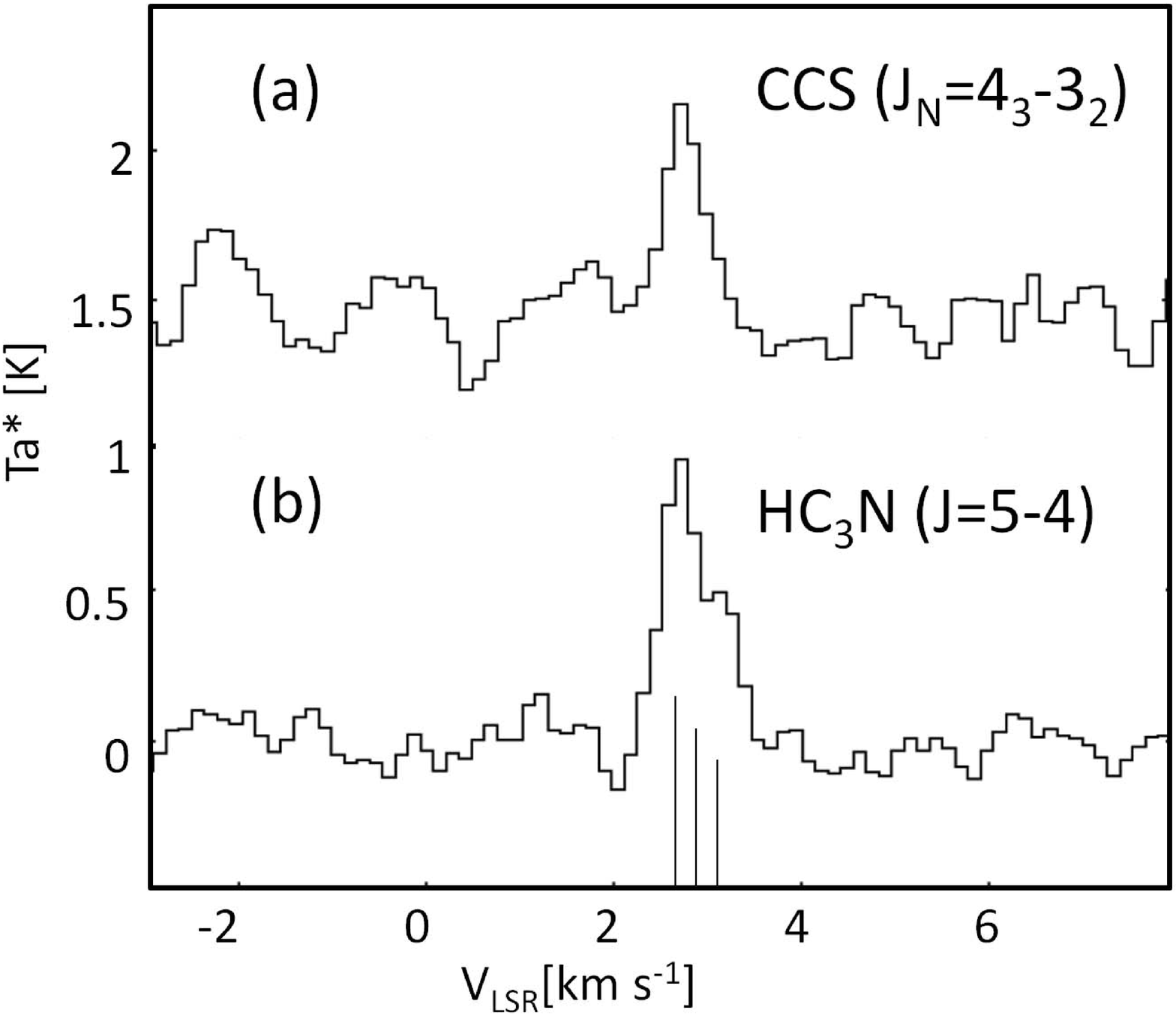}
\caption{Sample spectra of the (a) CCS ($J_N$=$4_3$-$3_2$) at (${\Delta}{\alpha}$, ${\Delta}{\delta}$)=(-40'',~-120'') and (b) HC$_3$N ($J$=5-4) at (${\Delta}{\alpha}$, ${\Delta}{\delta}$)=(0'',~-40''). The HC$_3$N line has a broader linewidth due to partially resolved three main (${\Delta}F$=1) hyperfine structure lines, whose theoretical locations and intensity pattern are shown at the bottom of the figure. Note that two weaker (${\Delta}F$=0) hyperfine structure lines are located outside of this figure.
{\bf Note that the temperature scale for the CCS line is offset by 1.5~K. }
}
\label{fig:samples}
\end{figure}

Figure~\ref{fig:samples} shows examples of the observed spectra for the individual molecules.
The top panel shows the CCS spectrum at its peak ($\Delta\alpha$, $\Delta\delta$) = (-40",~-120"), and the bottom corresponds to the HC$_3$N spectrum at its peak ($\Delta\alpha$, $\Delta\delta$) = (0",~-40"), respectively.
%
The highest antenna temperatures were 0.89~K for CCS and 0.95~K for HC$_3$N, respectively, in our mapping region.
Especially, in the Figure~\ref{fig:samples} (b), we could not observe the hyperfine structure (${\Delta}F$=0) lines of HC$_3$N, suggesting that the HC$_3$N lines are optically thin.

\begin{figure}
\plotone{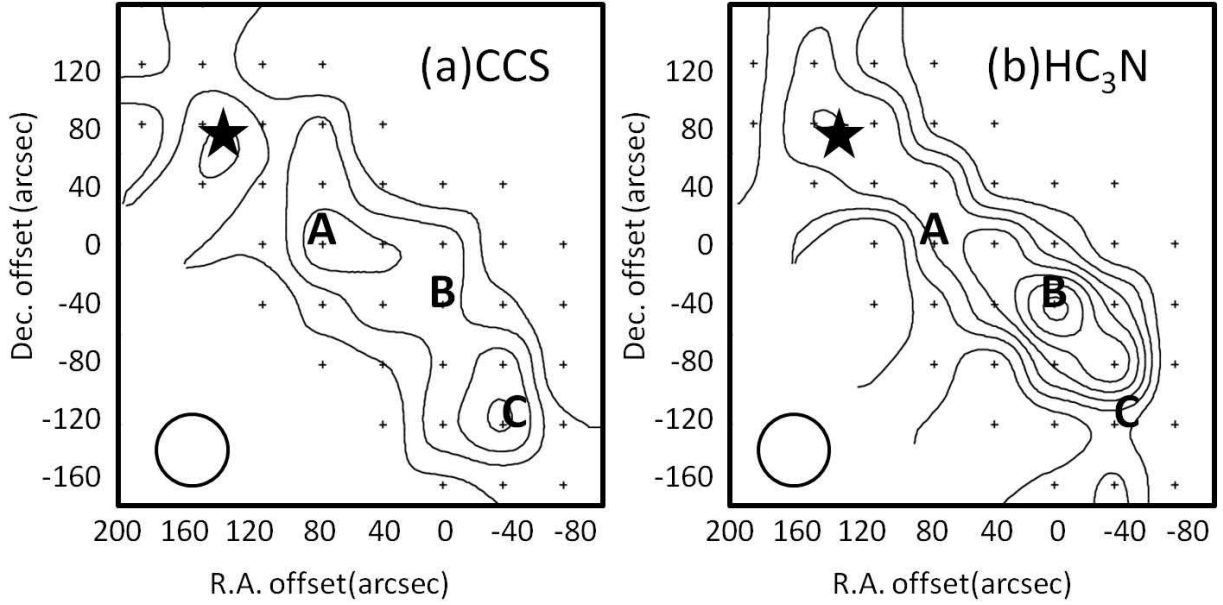}
\caption{
Integrated intensity maps of the observed molecules towards L1147.
The reference position is $\alpha _{2000}$=20$^{\rm h}$40$^{\rm m}$32$^{\rm s}$.0, $\delta _{2000}$=67$\degr$21'45".
The mapping positions are shown by the cross marks which trace a filamentary area seen in the dust continuum map in Figure~3.
The Spitzer point source (SSTc2d J204056.66+672304.9) is shown by the black star.
The beam size is shown at the lower left corner.
(a) CCS ($J_N$=$4_3$-$3_2$): the velocity range of integration is from 2.3 to 3.3~km~s$^{-1}$.
The contour interval is 0.07~K~km~s$^{-1}$, and the lowest contour level is 0.05~K~km~s$^{-1}$.
(b) HC$_3$N ($J$=5-4): the velocity range of integration is from 2.3 to 3.3~km~s$^{-1}$.
The contour interval is 0.07~K~km~s$^{-1}$, and the lowest contour level is 0.05~K~km~s$^{-1}$.
Three prominent cores are indicated as A--C.\label{fig:integ}}
\end{figure}

\begin{figure}
\plotone{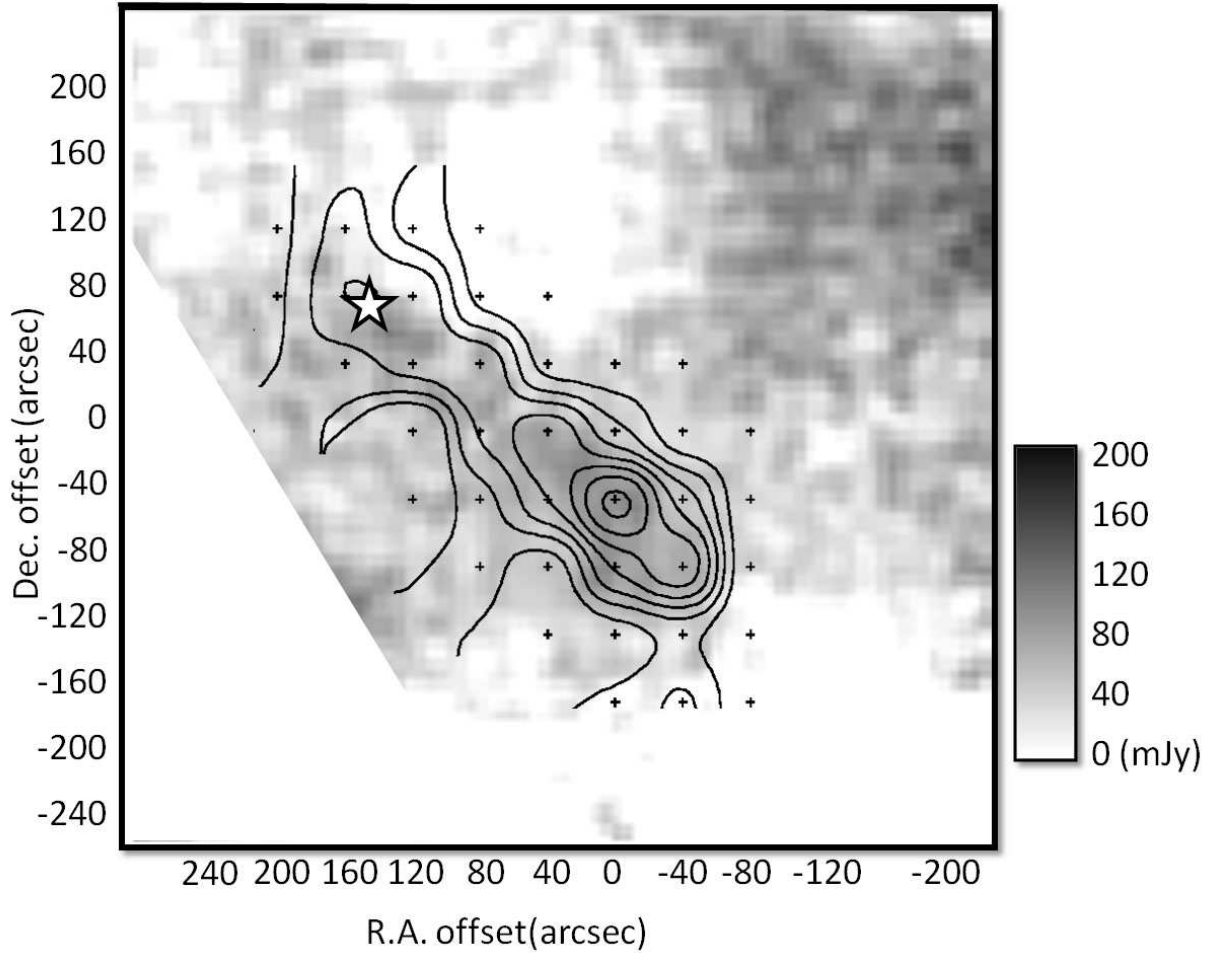}
\caption{
The dust continuum map obtained by SCUBA \citep{Kirk05}.
Dust continuum distribution is elongated toward NE--SW direction as shown in gray scale.
Integrated intensity map of HC$_3$N is superposed.
This trend matches well with CCS and HC$_3$N distribution.
Our observation positions are denoted by the ``+'' marks. 
\label{fig:dust_continuum}}
\end{figure}

Figures~\ref{fig:integ}(a) and \ref{fig:integ}(b) show integrated intensity maps in the CCS $J_N$=$4_3$-$3_2$ line and HC$_3$N $J$=5-4 line, respectively.
The velocity range of integration is from 2.3 to 3.3~km~s$^{-1}$.
We found that the CCS and HC$_3$N emissions are elongated in the NE--SW direction, which is consistent with the 850~$\micron$ map obtained by \cite{Kirk05} (see Figure~\ref{fig:dust_continuum}, 850~$\micron$ map from the SCUBA Legacy Catalogues \citep{Francesco08}). 
%
%
There are two CCS peaks and one HC$_3$N peak on the same filament: the two CCS peaks are located at (80",~0") and (-40",~-120"), and one HC$_3$N peak is found at (0",~-40").
%
%
Such the CCS distribution agrees well with the CCS $J_N$=$2_1$-$1_0$ map obtained by \cite{Kauffmann05}.
Hereafter we call these three cores as ``cores A, B and C'', from east to west, respectively.
The line parameters in these core are summarized in Table~\ref{table1}.
The HC$_3$N emission lines towards cores A and B show slightly broader linewidths due to blended hyperfine structure lines (see Figure~1(b)).

The dust continuum distribution, which would trace the H$_2$ density, is the strongest towards core B, followed by cores A and C.
H$_2$ column densities were derived from the dust continuum map by using the methodology described in \cite{Kauffmann08}, assuming dust opacity is 0.01~cm$^2$~g$^{-1}$. 
{\bf We obtained the H$_2$ column densities of $6.2\times10^{21}$~cm$^{-2}$
and $1.1\times10^{22}$~cm$^{-2}$ for cores A and B, respectively. 
The estimated error is $2.5\times10^{21}$~cm$^{-2}$ in the both cores, calculated from the typical noise fluctuation of dust continuum distribution in emission-free area.
\cite{Kauffmann08} reported an estimated H$_2$ column density to be $7.9{\pm}0.8 \times 10^{21}$~cm$^{-2}$, which is comparable to our estimated values.
However, for core C, we were able to derive a $3\sigma$ upper limit to the H$_2$ column density of $5.7\times10^{21}$~cm$^{-2}$ since the signal-to-noise ratio was not sufficient.}

The contours shown in Figures~\ref{fig:integ}(a) and \ref{fig:integ}(b) show that each core would have slightly larger size 
than our beam of 40".
Thus, their linear sizes would be around  0.07~pc or more, considering the distance of 325~pc.
In Figure~\ref{fig:integ}(b) the contours around cores B and C are interconnected.
This might mean that the two cores correspond to substructures within a single core.

The VeLLO source (SSTc2d J204056.66+672304.9) is shown as a black star in Figures~\ref{fig:integ}(a) and \ref{fig:integ}(b).
%
%
%
%
%
Prior to the mapping observations we expected to reveal a core in the vicinity of the VeLLO source since a protostellar core is often associated with an infrared source. However, we did not see a prominent core near the VeLLO source.

\begin{figure}
\plotone{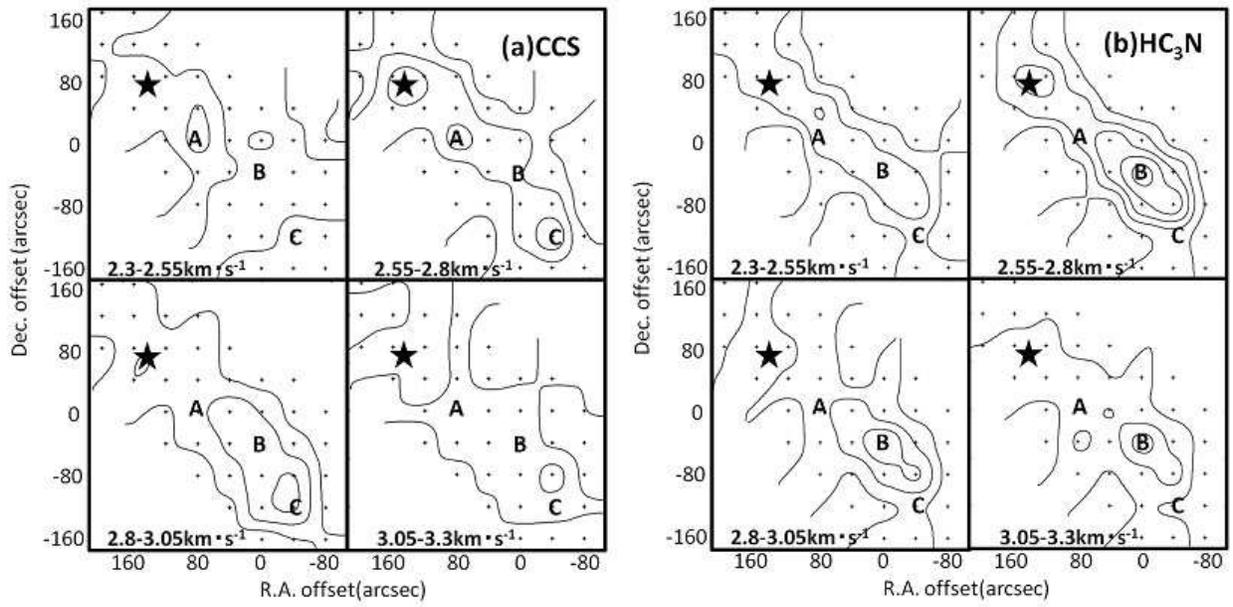}
\caption{
The channel maps of (a) CCS and (b) HC$_3$N.The contour interval is 0.04~K~km~s$^{-1}$ and the minimum contour value is 0.01~K~km~s$^{-1}$ for both species. 
The mapping region is the same as in Figure~\ref{fig:integ}.
\label{fig:ch}}
\end{figure}

Channel maps of CCS and HC$_3$N are shown in Figures~\ref{fig:ch}(a) and \ref{fig:ch}(b).
From Figure~\ref{fig:ch}(a) it is possible to mention that the two CCS cores have slightly different $V_{\rm LSR}$ values; the core C has the $V_{\rm LSR}$ value of 2.8--3.3~km~s$^{-1}$, while the core A has a slightly blue-shifted value. 
Further the core C appears as a more prominent core than the core A.
In Figure~\ref{fig:ch}(b) only the core B is visible, which has wider line width, 
and $V_{\rm LSR}$ is 2.3--3.5~km~s$^{-1}$ and is slightly extends along the NE--SW ridge. 
No major HC$_3$N peak corresponds to the cores A and C, however, it is seen that HC$_3$N emission exists at around the two CCS peaks.
Detailed descriptions for individual cores are summarized in Table~\ref{table1}.

In our mapping observations, we found that the three peaks are clearly interleaved to each other, 
contrary to the past observation results where CCS and HC$_3$N show similar 
spatial distributions \citep{Hirahara92, Tafalla06, Hirota02, Hirota04}.
For example, Figure 4 in Hirota et al.(2009) showed the integrated intensity maps of CCS 
and HC$_3$N towards seven dark clouds. The spatial distributions of CCS and HC$_3$N in six dark clouds
are very similar, including their peak positions, even if there is a slight difference in the distributions.
Only one exception is L1172D: there are two CCS peaks and a HC$_3$N is located between the CCS peaks.
It is well known that carbon-chain molecules are sometimes depleted in the central part of an infrared source and that their distributions show a central hole \citep{Velusamy95, Kuiper96, Ohashi99, Lai00, Lai03}.
The observed characteristics of the interleaved peaks seem to have different spatial distributions from the ``central-hole'' case mentioned above, and we will discuss later how such the interleaved distribution could be explained.

\subsection{Position-Velocity Maps}
\begin{figure}
\includegraphics[scale=.4]{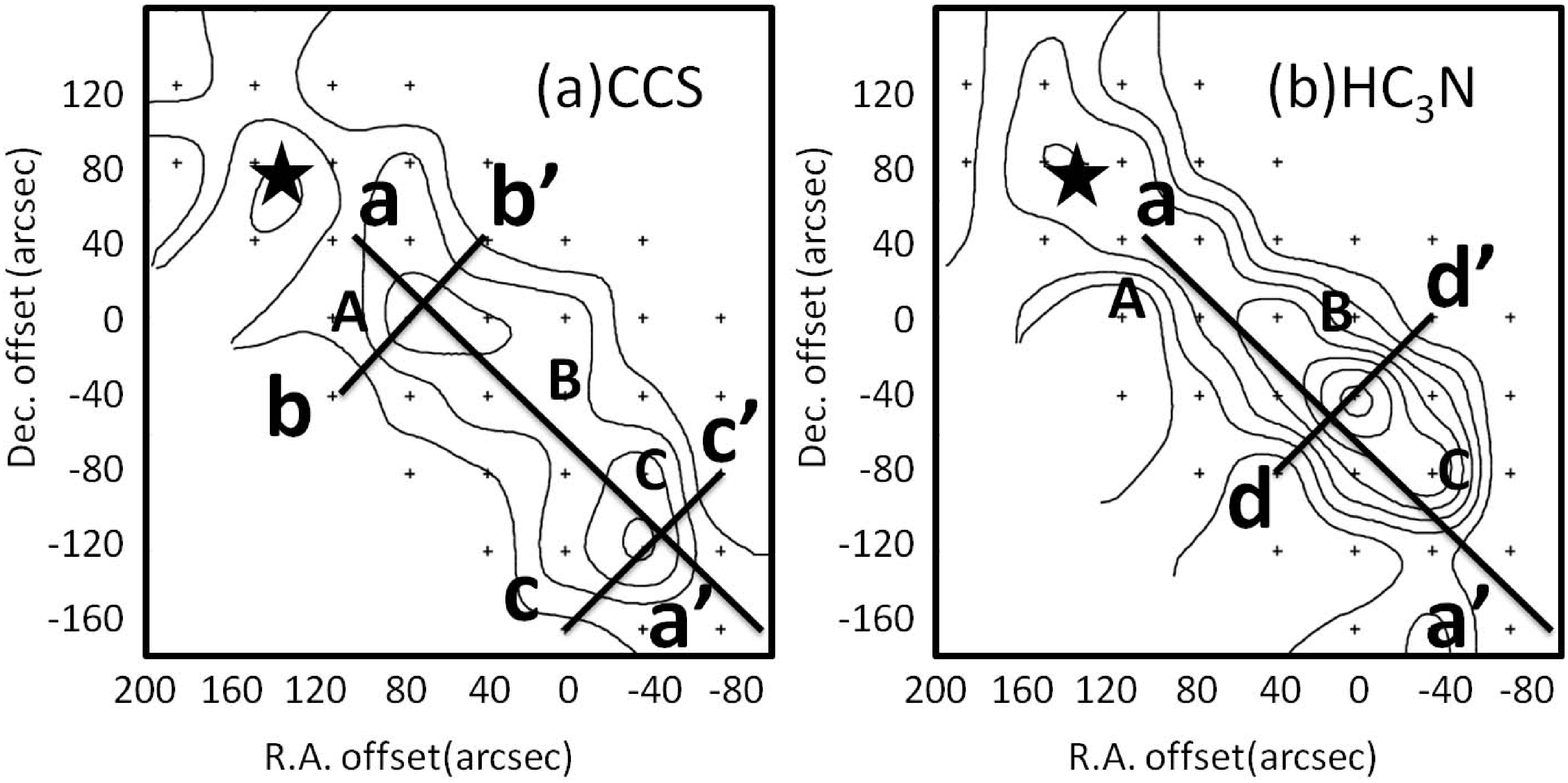}

\includegraphics[scale=.4]{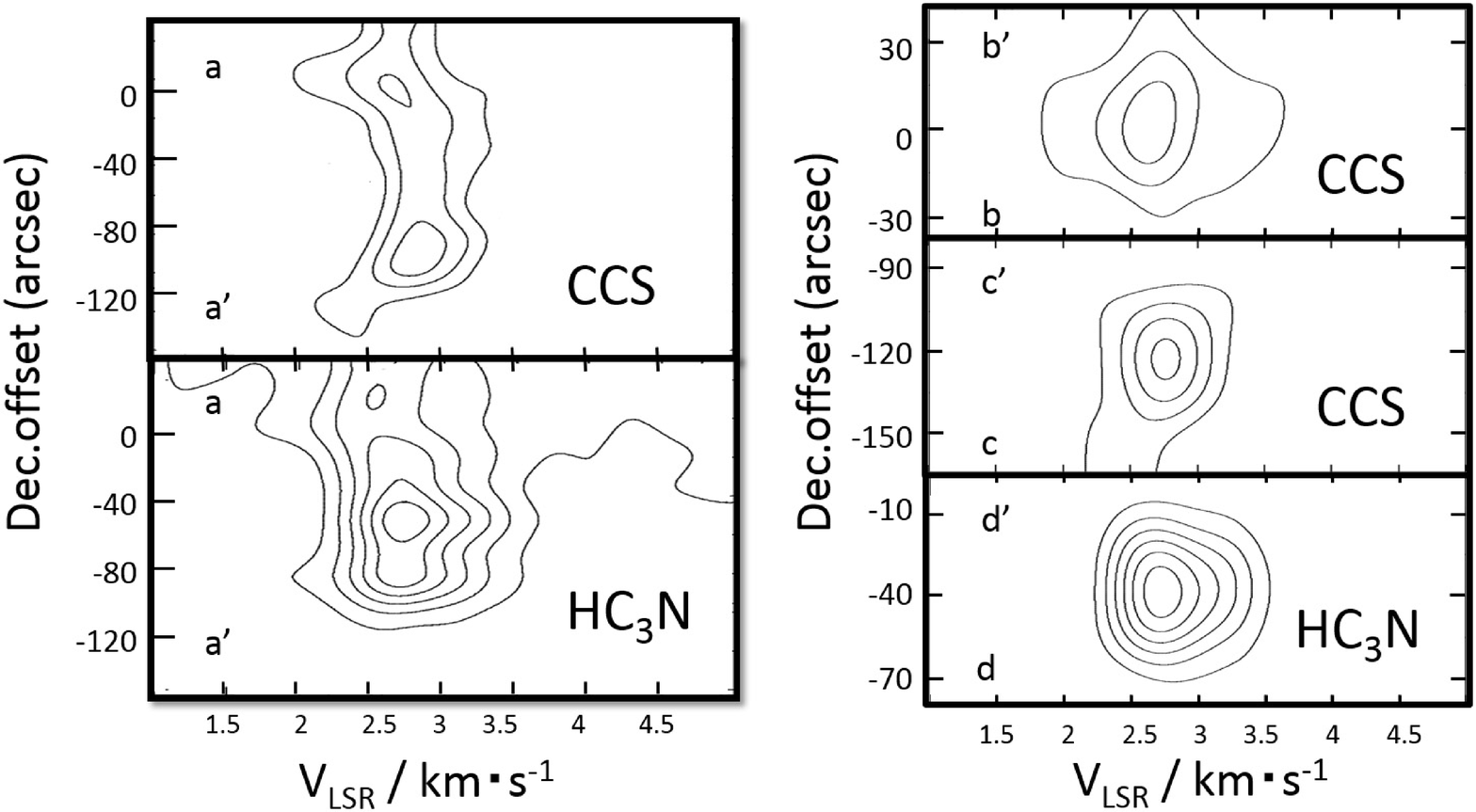}
\caption{
Position-velocity maps of the CCS $J_N$=$4_3-3_2$ and HC$_3$N $J=5-4$ lines. 
The cutting lines are shown in the upper integrated maps, as (a--a'), common cutting both in CCS and HC$_3$N maps,  (b--b'), (c--c'), and (d--d').
The contour interval is 0.1~K~km~s$^{-1}$, and the lowest contour level is 0.07~K~km~s$^{-1}$.
Three prominent cores are indicated as A--C, as is the case for Figure~\ref{fig:integ}.
Lower figures show position-velocity maps along each cutting line.
\label{fig:PV}}
\end{figure}

Figure~\ref{fig:PV}  shows position-velocity maps in L1147.
For CCS, we have put a cutting line (a--a') along the edge passing through the cores A and C. 
Additionally we have set two cutting lines perpendicular to the line a--a', passing through these cores.
We drew four lines in  Figure~\ref{fig:PV} (a) and (b), including three cores (b--b' and c--c').
Similarly for HC$_3$N, we have set a cutting line (d--d')  perpendicular to the line (a--a') and passing through the core B. 
%
%
%
%
The position velocity maps along the ridge (bottom left in Figure~\ref{fig:PV}) and perpendicular to the cores (bottom right in Figure~\ref{fig:PV}) do not show any sign of rotation.
%
%
This fact would be consistent with that the three cores are in an very early phase in their physical evolution.

\subsection{Molecular Column Densities}
We derived the column densities of CCS and HC$_3$N.
For the calculations, we used the following equations with assuming the local thermal equilibrium (LTE) condition and that the both lines are optically thin:
\begin{equation}
T^*_{\rm A}=\eta_{\rm mb}[J(T_{\mathrm{ex}})-J(T_{\mathrm{BB}})][1-\mathrm{exp}(-\tau )]
\end{equation}
where
\begin{equation}
J(T)=(h\nu /k)[\mathrm{exp}(h\nu /kT)-1]^{-1},
\end{equation}
\begin{equation}
N=\frac{3h}{8\pi^3}\frac{Q}{\mu^2S_{\rm ij}}\Delta v\frac{\mathrm{exp}(E_{\rm u}/kT_{\mathrm{ex}})}{\mathrm{exp}(h\nu/kT_{\mathrm{ex}})-1}\tau
\end{equation}

In these equations, $\eta_{\rm mb}$ denotes the main beam efficiency, $\nu$ the rest frequency of the line, $N$ the column density, $\Delta v$ the line width (FWHM), $Q$ the rotational partition function, $S_{\rm ij}$ the intrinsic line strength of the transition, $\mu$ the permanent dipole moment, $E_{\rm u}$ the upper state energy from the ground state rotational level. 
%
%
Since we observed a single line for each molecule, the excitation temperatures were fixed to 5~K for CCS and 6.5~K for HC$_3$N, which are consistent with our previous studies \citep[e.g.,][]{Suzuki92, Hirota09}.
%
A change of excitation temperature by 1~K makes a change of their column densities about 30$\%$. 

\begin{figure}
\includegraphics[scale=.4]{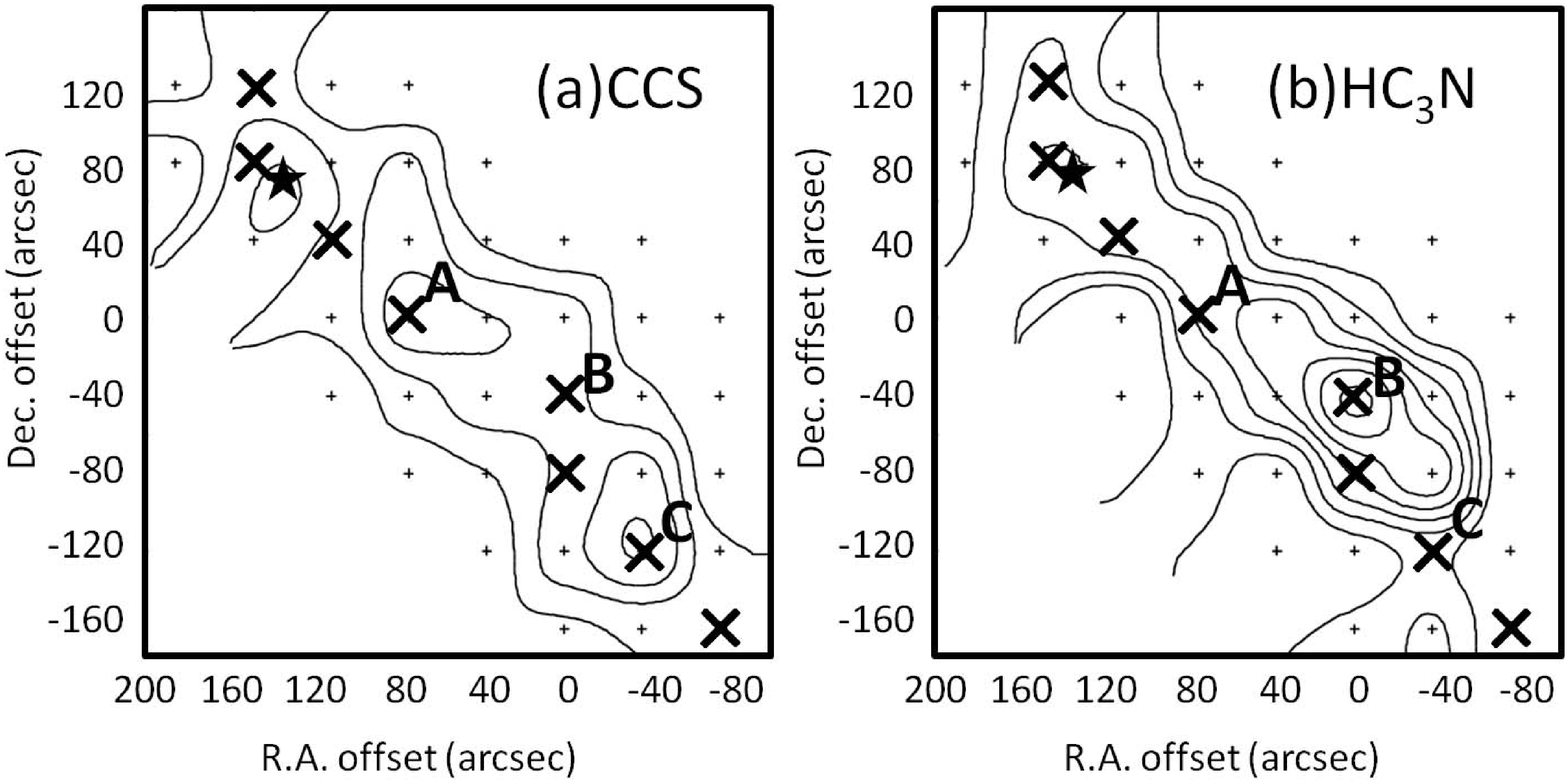}

\includegraphics[scale=.6]{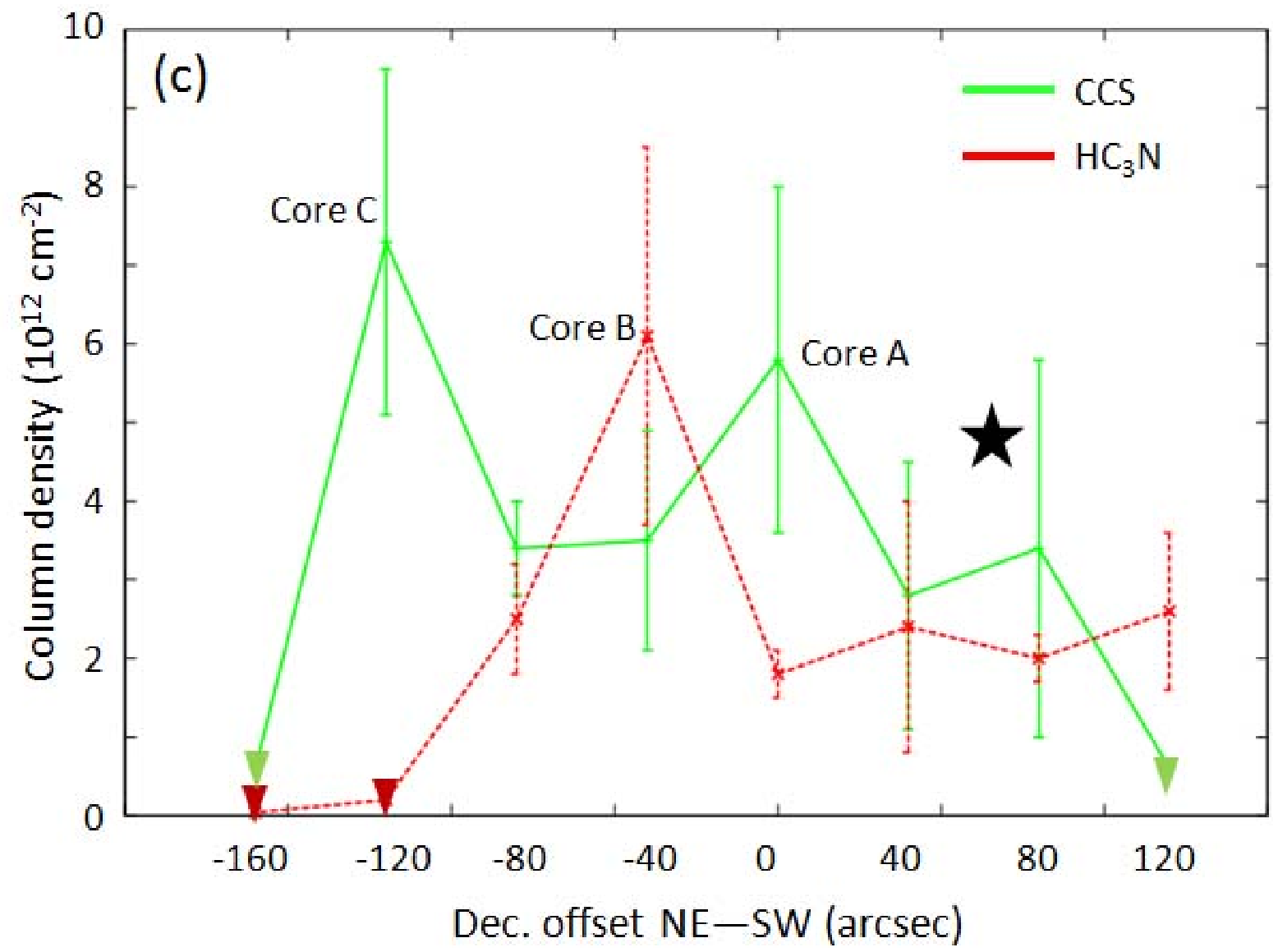}
\caption{
(a) and (b) We have chosen eight points along the ridge with different declination offsets;
(c) Distribution of column densities of CCS and HC$_3$N along the ridge.
The abscissa denotes the declination offset,
whose positions are shown in (a) and (b).
The position of the VeLLO source is shown by the black star. 
\label{fig:cdd}
}
\end{figure}

Figure~\ref{fig:cdd} shows derived column density distributions along the prominent NE--SW ridge seen in the integrated 
intensity maps (Figure~\ref{fig:integ}).
The column densities of the both molecules resulted in a range of 2--7$\times10^{12}$~cm$^{-2}$, which is consistent with the values reported by \cite{Hirota09}.
We have found that the derived column densities of CCS (3--7$\times10^{12}$~cm$^{-2}$) are less by about one order of magnitude than those in other known CCPRs, e.g., $66\times10^{12}$~cm$^{-2}$ for TMC1(CP) \citep{Suzuki92} 
and $53\times10^{12}$~cm$^{-2}$ for L492 \citep{Hirota09}. 

%
%
The three peaks (cores A, B and C) are also seen in the column density distributions along the NE--SW ridge, 
however, we have found that the column density ratio between CCS and HC$_3$N (hereafter called as ``CCS/HC$_3$N'' ratio) varies dramatically. 
 We summarized in Table~\ref{table2} their column densities, fractional abundances relative to H$_2$, and ``CCS/HC$_3$N'' ratios.
The errors of column densities are calculated by using the r.m.s noise in each observation.
The upper limit to the column density of HC$_3$N in core C was calculated using the $3\sigma$ r.m.s noise at this position.
We have taken the estimated error of H$_2$ into account when calculating the fractional abundance values.
{\bf Note that the dust emission in core C is weaker than the noise level.
Therefore we used 3~~$\sigma$ level of r.m.s noise for H$_2$ density to calculate the upper limit of the fractional abundances of CCS.
}
In core A, CCS has higher abundance than that of HC$_3$N. 
On the other hand this trend is reversed in core B, as revealed in their column density ratio.
Especially, core C, where HC$_3$N was not seen, has an extremely high CCS abundance; we are able to indicate only the lower limit to the abundance ratio.
{\bf Since both the H$_2$ and HC$_3$N column densities are given as lower limits, we were not able to estimate the fractional abundance of HC$_3$N toward core C.}

\section{Discussion}
In this section we will discuss if L1147 can be regarded as a CCPR, and how we could reconcile the displacement of emission peaks and the variation in the CCS-to-HC$_3$N ratio among the prominent cores.

%

\subsection{NH$_3$ Data in the L1147 Filament}
The first issue that we want to clarify is if L1147 is a CCPR as a whole. 
In this regard, it would be important to know how strong or weak the NH$_3$ line is.
\cite{Hirota09} reported the NH$_3$ (1,1) data only 
towards the (0",~-80") position by using the Nobeyama 45m radio telescope, which shows very weak emission (the antenna temperature of 80~mK). 
\cite{Hirota13} {\bf observed the NH$_3$ (1,1) line towards the IRS position (the VeLLO source), however, the line was too faint to detect with an r.m.s. noise level of 30~mK.}
Therefore it is possible to calculate an upper limit to the column density of NH$_3$ towards the VeLLO source with the same method used by \cite{Hirota09}, which is $9\times10^{12}$~cm$^{-2}$. This value is comparable with that reported by
\citep{Hirota09} ($8\times10^{12}$~cm$^{-2}$).
Since the beam size is large ($\sim$80~arcseconds), the two beams covered the most part of the mapping area in the 
CCS and the HC$_3$N lines.
Thus we may conclude that the column density of NH$_3$ in the L1147 filament is less than $9\times10^{12}$~cm$^{-2}$ as a whole.
When we use this upper limit to the NH$_3$ column density together with the column densities of CCS tabulated in Table~\ref{table2}, the ``NH$_3$/CCS'' ratio is found in a range between 0.95 and 4.3.
This leads us to confirm that the entire L1147 filament is a CCPR.

\subsection{Variation of ``CCS/HC$_3$N" Ratios}
\begin{figure}
\includegraphics[scale=0.6]{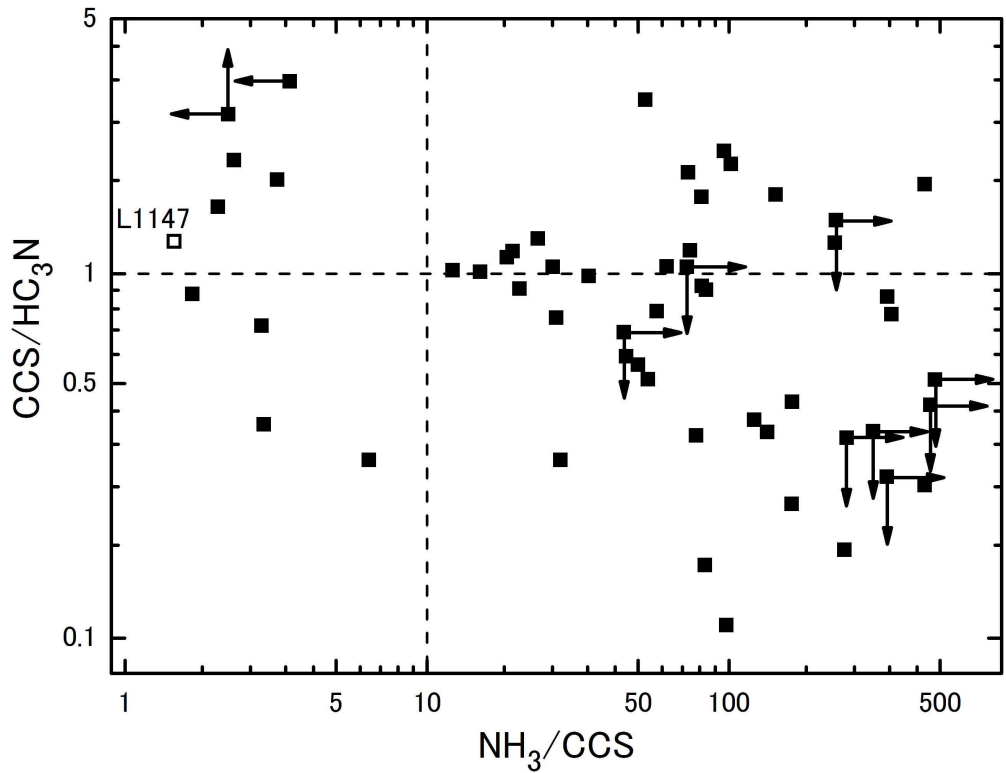}
\caption{
``CCS/HC$_3$N'' ratio as a function of ``NH$_3$/CCS'' ratio in various clouds.
The data points were taken from \cite{Suzuki92} and \cite{Hirota09} where the CCS, HC$_3$N and NH$_3$
data are available. The arrows indicate upper or lower limits to the ratios. The data point corresponding to L1147 is shown in the open square.
\label{fig:ratio}
}
\end{figure}

It is well known that abundances of carbon-chain molecules increase rapidly then decrease as the cloud evolves \citep{Suzuki92}.
Although \cite{Suzuki92} reported observation results for CCS, HC$_3$N, HC$_5$N and NH$_3$, they did not compare abundances of CCS and HC$_3$N based on their gas-phase chemical evolution model.
Further, at the best of our knowledge, no one has conducted such comparisons, especially in the early phase of cloud evolution.
The ``CCS/HC$_3$N" ratio is expected to be uniform since cores formed in a single filament would have similar evolutionary phase. 
However, we have found that the three cores in the L1147 filament show different the variation of ``CCS/HC$_3$N" ratios.
This would suggest that the three cores are in different evolutionary phases.
Then can we see such variations of the ``CCS/HC$_3$N" ratio for other dark clouds ?
Figure~\ref{fig:ratio} shows a {\bf cloud-to-cloud variation} of the ``CCS/HC$_3$N" ratio as a function of the ``NH$_3$/CCS" ratio.
The data were taken from \cite{Suzuki92} and \cite{Hirota09} where the CCS, HC$_3$N and NH$_3$ data are available.
Since it is well known that the ``NH$_3$/CCS" ratio is a good indicator of chemical evolution of a cloud, this plot represents
a time-evolution of the ``CCS/HC$_3$N" ratio along with the chemical evolution.
A {\bf weak} trend can be seen from this figure that a majority of cores in the  CCPRs and their candidates (the ``NH$_3$/CCS" ratio is less than 10) have higher ``CCS/HC$_3$N" ratio  whereas a majority of  evolved cores have lower ``CCS/HC$_3$N" ratio.

In order to understand difference of the three cores in L1147 as well as the trend of the ``CCS/HC$_3$N" ratios, 
it would be a methodology to utilize chemical reaction network simulations for understanding chemical evolution 
of early phase molecular clouds. 

\subsection{Chemical Evolution Model}
We have conducted chemical evolution simulations based on a latest theoretical model that incorporates surface reactions on the dust particles.

%
%
For chemical reactions and their rate coefficients, we used the KIDA (KInetic Database for Astrochemistry) dataset \citep{Wakelam12}, which is a collection of measured and theoretical values published in literatures.
The simulation code was also provided by \cite{Wakelam12} , and includes 684 molecules and 8317 reactions.
The simulation code numerically solves the chemical kinetics equations, shown below, that describes the formation and destruction of molecules.

\begin{equation}
\frac{\mathrm{d}n_i}{\mathrm{d}t}=\sum_{l,m}k_{lm}n_ln_m-n_i\sum_{i\neq l}k_ln_l+k_{i}^{\mathrm{des}}n_i^s-k_i^{\mathrm{acc}}n_i
\end{equation}
\begin{equation}
\frac{\mathrm{d}n_i^s}{\mathrm{d}t}=\sum_{l,m}k_{lm}^sn_l^sn_m^s-n_i^s\sum_{i\neq l}k_l^sn_l^s+k_{i}^{\mathrm{des}}n_i^s-k_i^{\mathrm{acc}}n_i
\end{equation}
where $n_i$ and $n_i^s$ are the gas-phase and surface concentrations of the $i$-th species (cm$^{-3}$), $k_{lm}$ and $k_l$ are the gas-phase reaction rates (in units of s$^{-1}$ for the first-order kinetics and cm$^3$~s$^{-1}$ for the second-order kinetics), $k^{acc}_i$ and $k^{des}_i$ denote the accretion and desorption rates (s$^{-1}$), and $k_{lm}^s$ and $k_l^s$ are the surface reaction rates (cm$^3$~s$^{-1}$).
It should be noted that this model assumes a constant number density of hydrogen atom and other physical parameters throughout a simulation. 
The gas kinetic temperature was fixed to 10~K, a typical value for a dark cloud, and the visual extinction, $A_{\rm v}$, was fixed to 10 magnitudes.
As the initial elemental abundances (see Table~\ref{table3}, we adopted the EA2  set used by \cite{Wakelam08}.

\begin{figure}
\includegraphics[scale=.25]{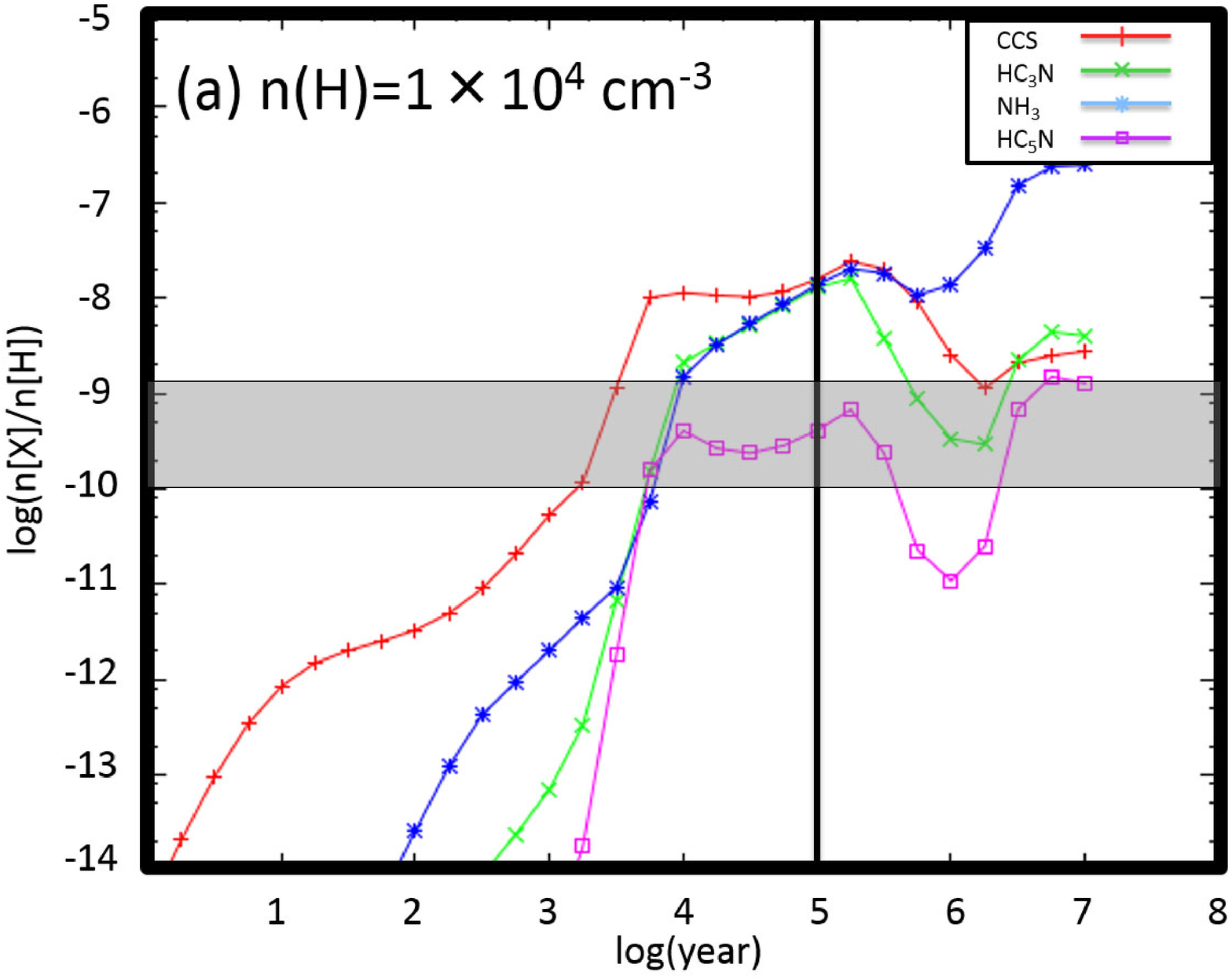}
\includegraphics[scale=.25]{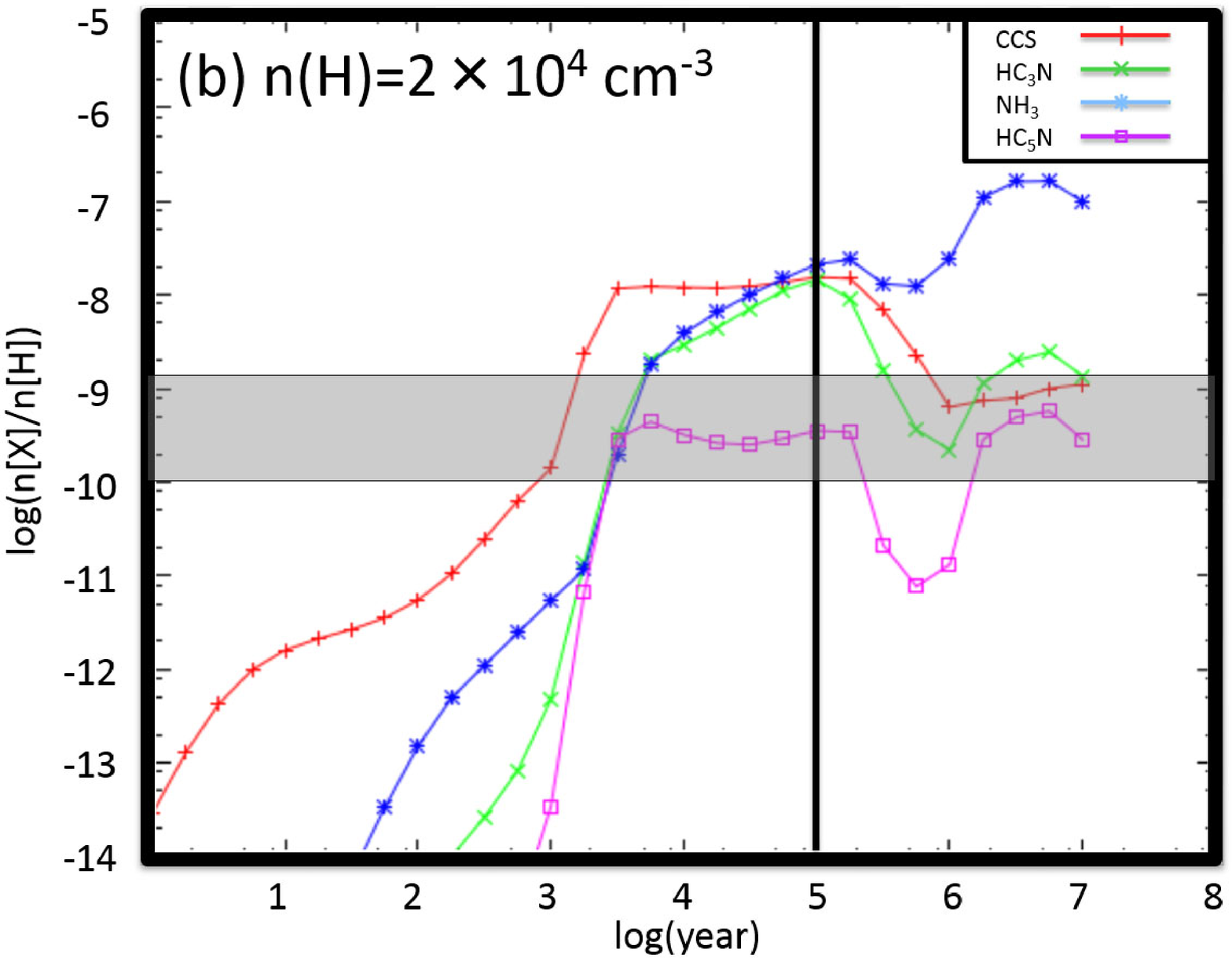}

\includegraphics[scale=.25]{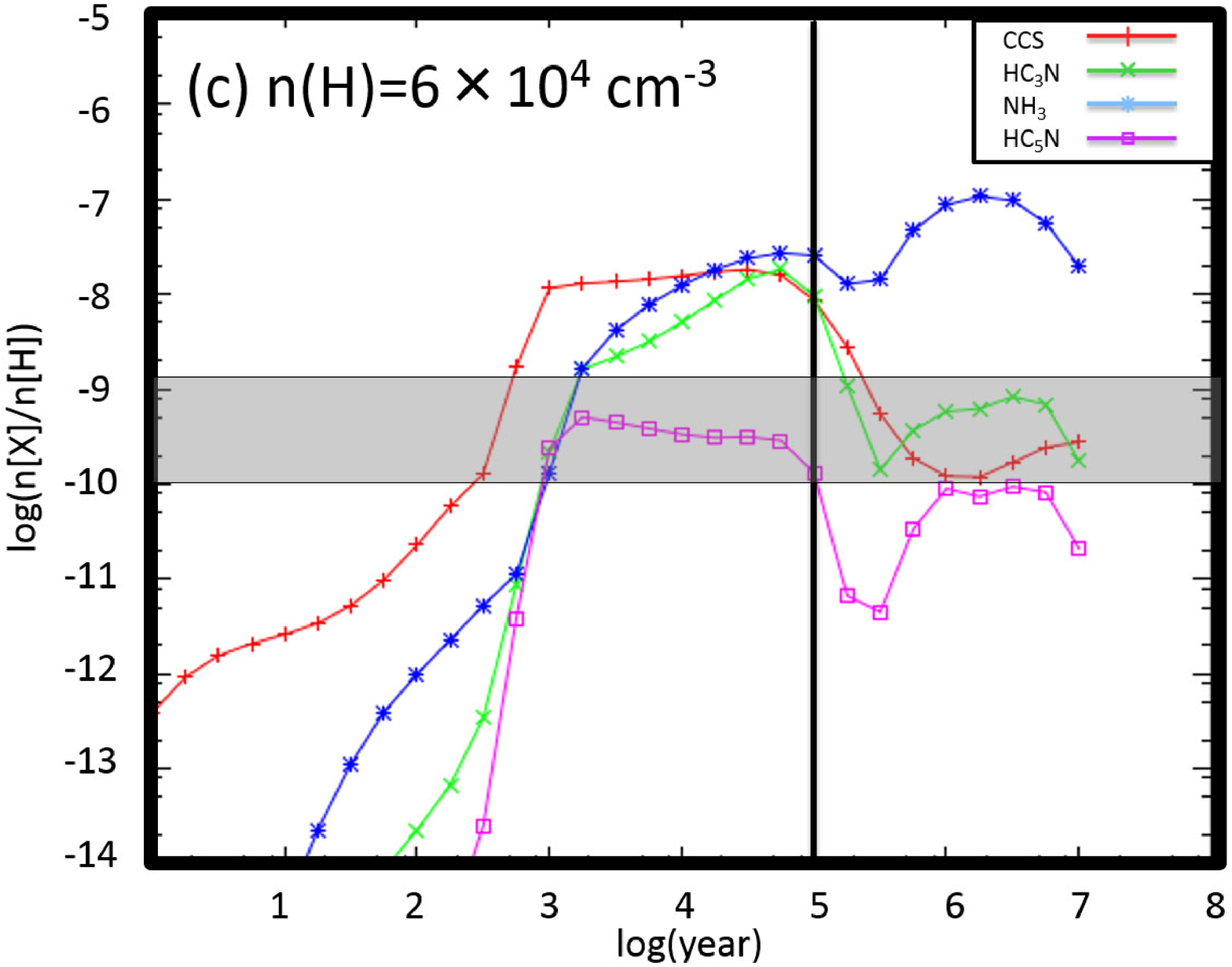}
\caption{
Examples of simulation results for different hydrogen number densities: (a) $1\times10^4$~cm$^{-3}$, 
(b) $2\times10^4$~cm$^{-3}$, and (c) $6\times10^4$~cm$^{-3}$.
Vertical line denotes $1\times10^5$ years, a typical age of a dark cloud core.
The column density range for CCS and HC$_3$N derived in our observation is shown in the gray region. 
\label{fig:simsample}
}
\end{figure}

Figure~\ref{fig:simsample} shows examples of our simulation results for the H density of 1$\times$10$^4$, 2$\times$10$^4$, 
and 6$\times$10$^4$~cm$^{-3}$.
The gray region corresponds to the observed column density range for CCS and HC$_3$N.
As can be seen from Figure~\ref{fig:simsample}, CCS is produced rapidly in the early phase (less than 10$^5$ years) of the cloud evolution and decrease afterward.
On the other hand, N-bearing species (HC$_3$N, HC$_5$N and NH$_3$) increase gradually.
Although HC$_3$N and HC$_5$N decrease along with the decrease of CCS, NH$_3$ increases monotonically in the whole evolution.
Such a behavior is similar to that reported and discussed in \cite{Suzuki92}.

Such behavior of their abundance is well connected with the formation process.
The formation process of CCS was discussed in \cite{Suzuki92}; the neutral carbon  (C) would play an important role for the
production of CCS.
%
%
As a cloud evolves, C$^+$ is converted into C, and finally into CO in the later phase.
CCS production is efficiently produced in the early phase of cloud evolution, and its fractional abundance is highest around 10$^4$-10$^5$ years, where the C atom shows its highest abundance among C$^+$, C and CO.

HC$_3$N, another carbon-chain molecule, also reaches to the maximum in the same time range.
Regarding its formation process, it was suggested that a neutral-neutral reaction between C$_2$H$_2$ and CN would be important \citep{Fukuzawa98}, which has no reaction barrier.
C$_2$H$_2$ is produced in relatively early phase of cloud evolution, while CN formation would longer time.
CN can be formed via CH + N$^+$ and/or NH + C.
As described in the formation of NH$_3$ below, these reactions are related to the ionization of N atom, which takes longer time.
As a result, HC$_3$N starts to increase a bit later than CCS in our simulations.
Both of CCS and HC$_3$N abundances decrease dramatically in the later phase of cloud evolution, because most of C atoms 
are converted into CO, and the formation of carbon-chain molecules is no longer efficient.

On the other hands, the fractional abundance of NH$_3$ shows very different behavior.
Laboratory experiments showed solid-phase reaction to form NH$_3$ would not be efficient, 
and most of NH$_3$ would be formed in the gas-phase \citep{Hiraoka06}.
In our simulation, it is assumed that only atomic nitrogen exist initially.
The formation of NH$_3$ would be initiated by the ionization of N by a cosmic ray particle (CRP).
Then sequential reactions with H$_2$ \citep{Herbst73} form NH$_4^+$.
Finally, NH$_3$ is produced through recombination of NH$_4^+$ with electron.
We summarize a series of reactions together with their rate coefficients $k$ at 10 K in the brackets.
%

N + CRP $\rightarrow$ N$^+$ + e$^-$ ($k$=$2.9\times10^{-17}$~cm$^3$~s$^{-1}$) (1)

N$^+$ + H$_2$ $\rightarrow$ NH$^+$ + H ($k$=$6.3\times10^{-12}$~cm$^3$~s$^{-1}$) (2)

NH$^+$ + H$_2$ $\rightarrow$ NH$^+_2$ + H ($k$=$1.0\times10^{-9}$~cm$^3$~s$^{-1}$) (3)

NH$^+_2$ + H$_2$ $\rightarrow$ NH$^+_3$ + H ($k$=$1.2\times10^{-10}$~cm$^3$~s$^{-1}$) (4)

NH$^+_3$ + H$_2$ $\rightarrow$ NH$^+_4$ + H ($k$=$1.2\times10^{-12}$~cm$^3$~s$^{-1}$) (5)

NH$^+_4$ + e$^-$ $\rightarrow$ NH$_3$ + H ($k$=$5.8\times10^{-6}$~cm$^3$~s$^{-1}$) (6)

Reaction (2) has a small rate coefficient since it has an activation energy of about 42~K. 
Although reaction (5) has a ``negative'' activation energy of about -36~K, the prefactor is so small ($3.36\times10^{-14}$ cm$^3$~s$^{-1}$).
Such small rate coefficients, including the ionization of N atom by a cosmic ray, would result in long time in forming NH$_3$.
%
%
%
%
We drew a vertical line at $10^5$ years in Figure~\ref{fig:simsample}, corresponding to a typical age for dark clouds.
The observed column densities of CCS and HC$_3$N agree with the simulation results within a factor of 10 (X$_{\rm obs}$/10 $\leq$ X$_{\rm model}$ $\leq$ 10X$_{\rm obs}$) where X$_{\rm obs}$ is the observed abundance and  X$_{\rm model}$ is the model abundance.
%
The same criterion is also used by experts in chemical reaction network simulations, \cite[e.g.,][]{Wakelam08}.
We find that our simulated column densities tend to be always higher than our observation results, and it would not be suitable to compare directly by means of individual column densities.
However, we found that the ratio of CCS to HC$_3$N was roughly matched with the past observations of CCPRs.
Thus, in the following section, we will use the ratio of CCS to HC$_3$N when comparing our observed results and simulated ones.
Furthermore, we can ignore the uncertainty of hydrogen column density by using their ratios.

\subsection{Comparison between Observational and Simulation Results}
We simulated chemical evolution under different densities from 10$^4$ to 10$^6$~cm$^{-3}$, and compared ``CCS/HC$_3$N" ratio with our observation between $3.16\times10^4$ and $1.0\times10^7$ years.
Some examples of simulation under different densities are shown in Figures~\ref{fig:simsample}(b) and (c).
It is seen that chemical evolution would be accelerated in physically evolved denser cores due to larger number of reactants per unit volume.

\begin{figure}
\plottwo{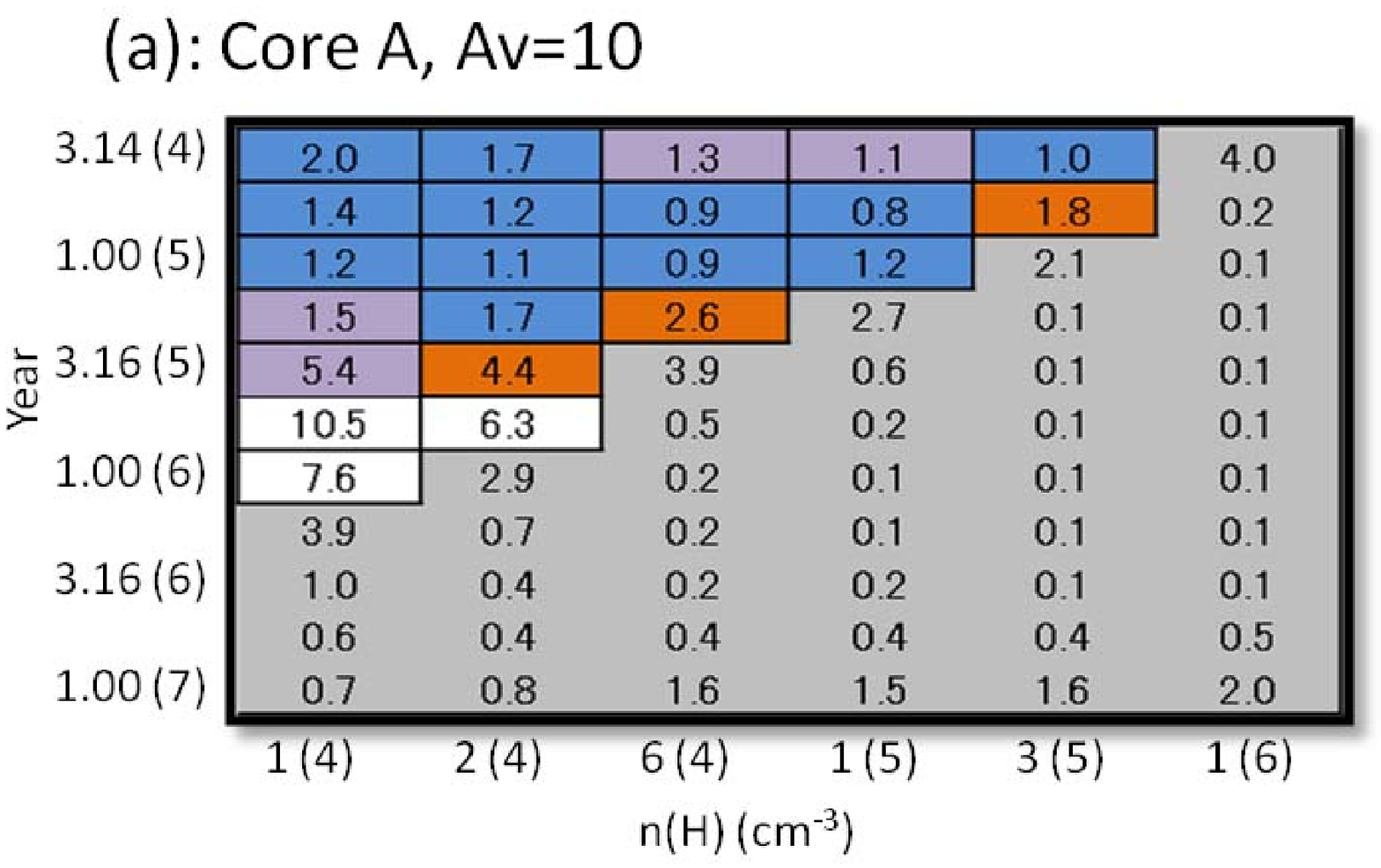}{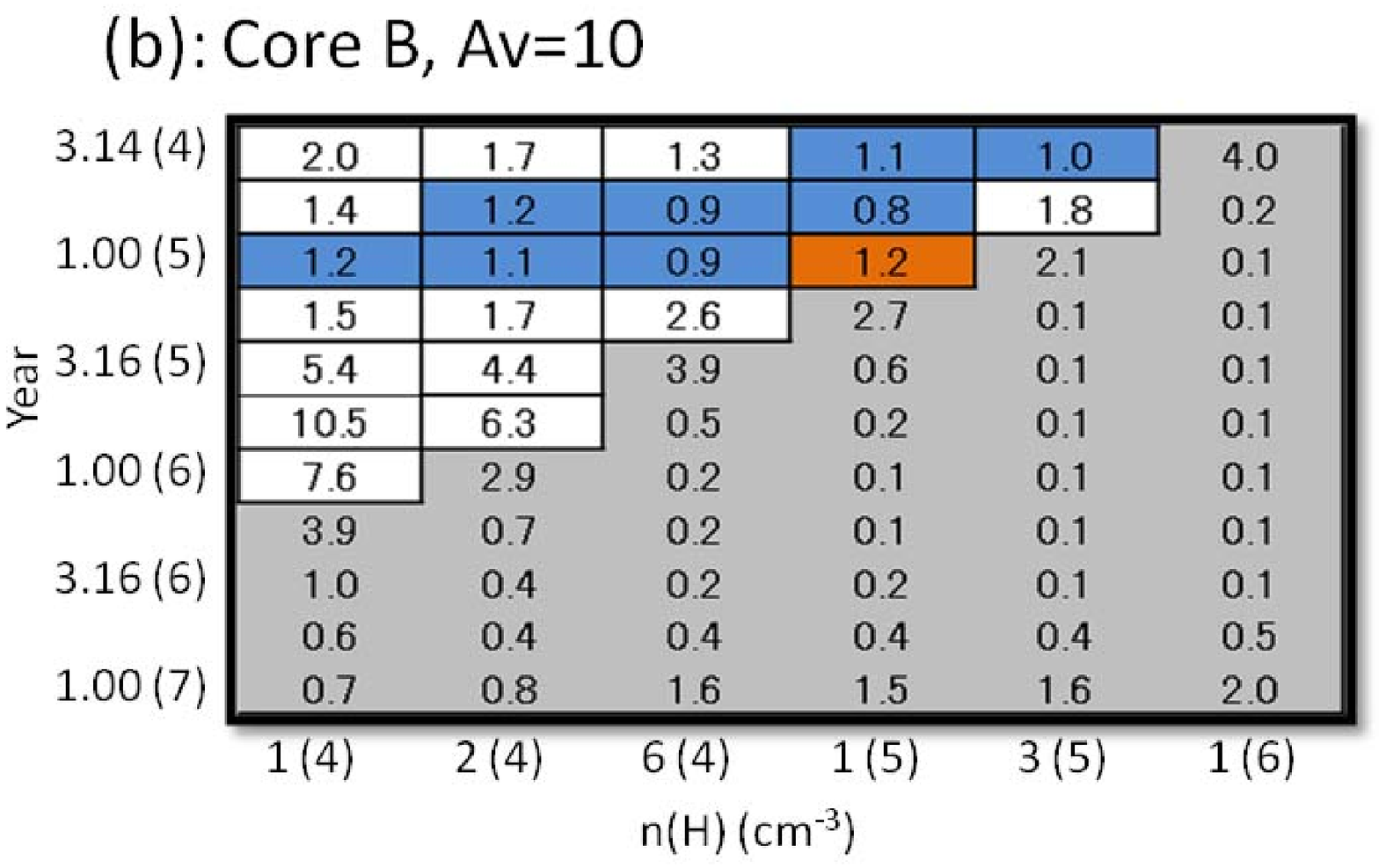}
\plottwo{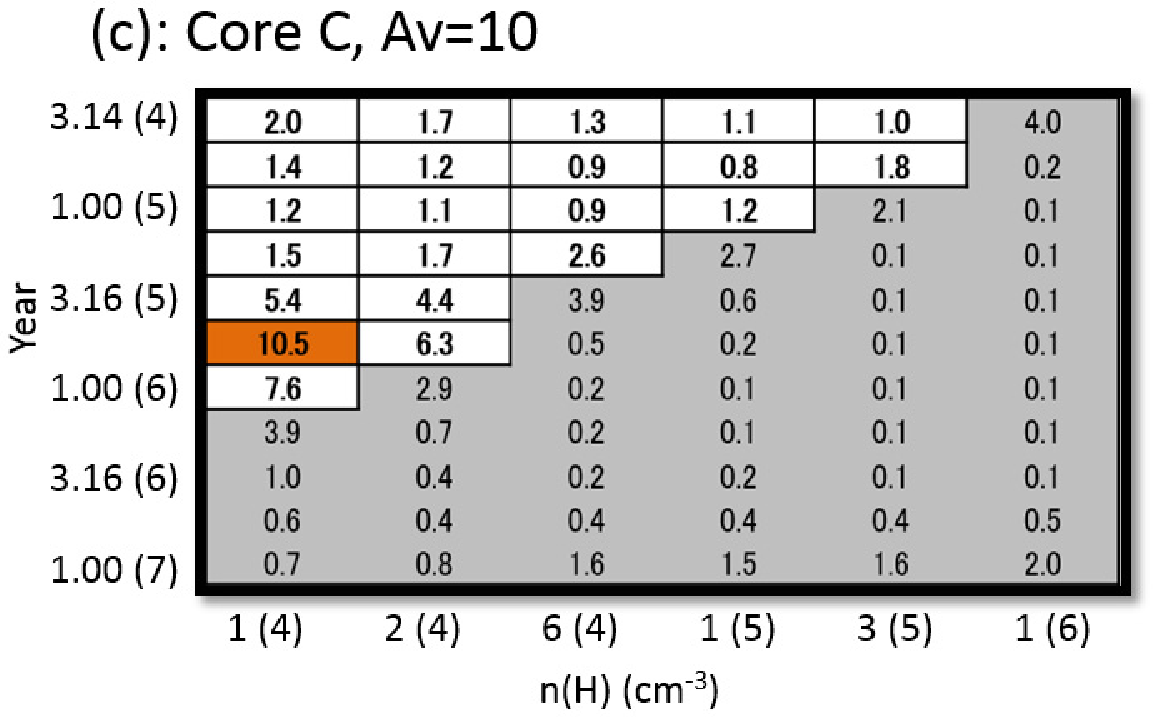}{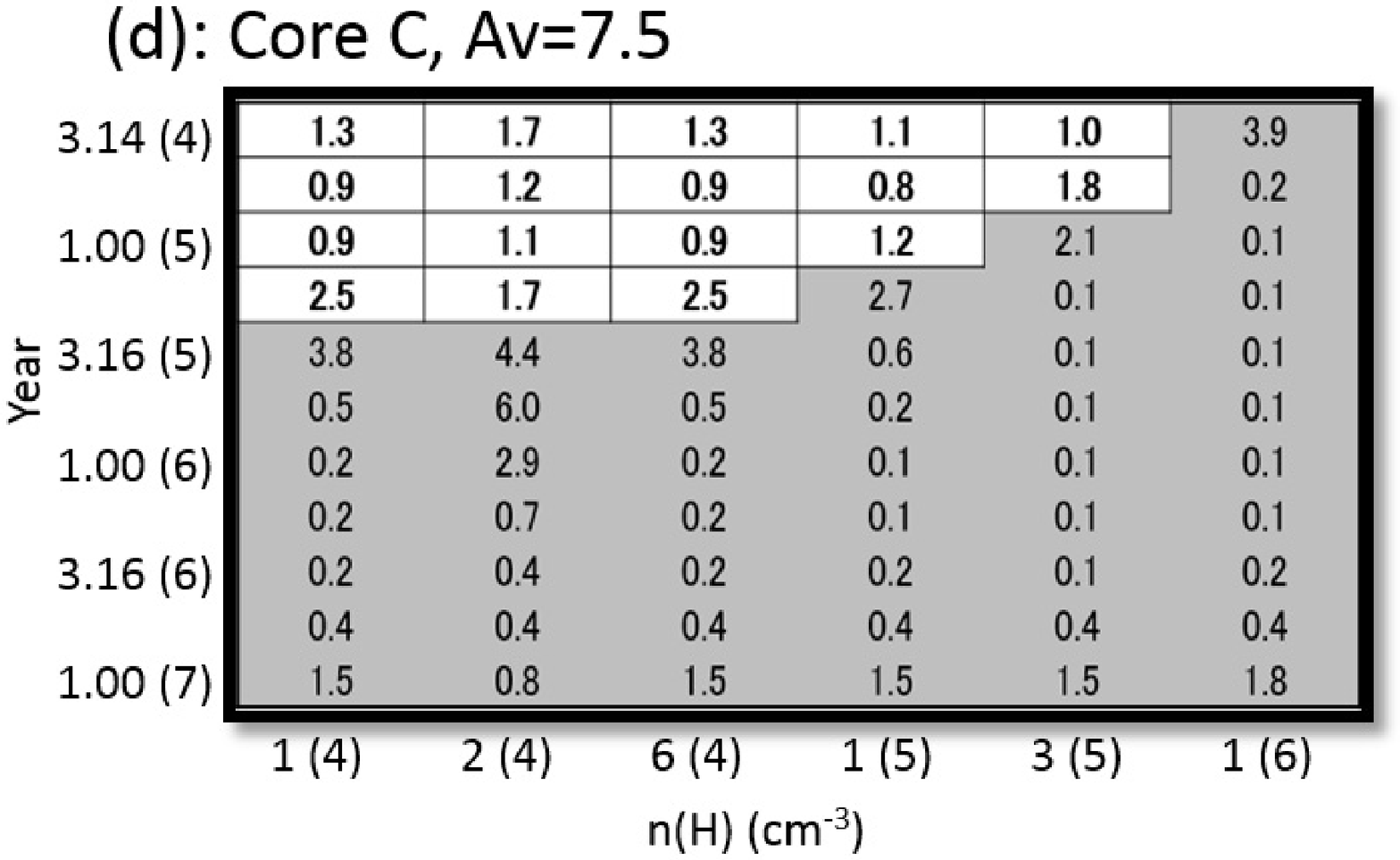}
\includegraphics[scale=1.0]{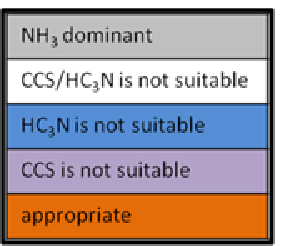}
\caption{
``CCS/HC$_3$N'' ratio derived in our simulation for different times and hydrogen number densities.
In the figures, time and density are represented in $\alpha$($\beta$) which means ${\alpha}{\times}$10$^{\beta}$.
We classify their values using different colors.
Grey regions corresponds to NH$_3$ dominant phase, white ones show unsuitable ``CCS/HC$_3$N'' ratios, 
and blue and purple ones show too much HC$_3$N and CCS abundance.
{\bf For Core C, where the fractional abundance of HC$_3$N was not determined, we took into account fractional abundances of NH$_3$ and CCS and the ``CCS/HC$_3$N'' only.}
Appropriate ratios and column densities are achieved in the orange region. 
(a), (b), and (c) represent comparison with cores A, B, and C, respectively.
Visual extinction is fixed to be 10 magnitude in (a)-(c).
For (d) we used the visual extinction to 7.5 magnitude, since it corresponds to the core C 
which would be a lower gas density region at an edge of the L1147 filament.
\label{fig:sim}
}
\end{figure}

Figures~\ref{fig:sim}(a), (b), and (c) show the simulated ``CCS/HC$_3$N'' ratios derived in different densities and ages.
Each figure has the same ratio value for the same age and the density, but we prepared three ones to compare simulation with the individual three cores since they have the different fractional abundance of CCS and HC$_3$N, and the ``CCS/HC$_3$N'' ratio.
We compared Figure~\ref{fig:sim} (a) with core A, (b) with core B, and (c) with core C.
Now we classify them in terms of the observed ratios.
As we expected, NH$_3$ becomes dominant in the later phase compared with carbon-chain molecules.
Since NH$_3$ is so weak in L1147, such an evolutionary stage would not be appropriate for L1147.
Thus we excluded them in our comparison (in Figure~\ref{fig:sim} the regions are shown in grey).
Some regions painted in white are not suitable in terms of the ``CCS/HC$_3$N'' ratios.
We also excluded regions where observed fractional abundances of CCS and HC$_3$N are not within a factor of 10 (the regions are shown in purple and blue).
{\bf For Core~C, we did not use the fractional abundance of HC$_3$N as criteria of comparison since we were only able to set upper limits to the column density of HC$_3$N and H$_2$.}
Then we found that the observed ``CCS/HC$_3$N'' ratios in cores A, B and C can be reproduced in orange regions 
between $6\times10^4$--$1\times10^6$ years, typical for dark clouds in Figures~\ref{fig:sim} (a)--(c).
All appropriate regions are those just before NH$_3$ becomes dominant.
In realistic dark clouds, physical evolution would be accelerated along with the increase of gas density.
%

\subsection{Evolutionary Phases in the Cores}
In this last subsection we would like to discuss if the three cores were formed and evolved in a similar way.

Figures~\ref{fig:sim} (a)--(c) were made assuming the visual extinction of 10~magnitude.
Core C is located in the edge of the L1147 filament where the gas density could be lower than those in cores A and B, and could have a lower visual extinction.
Figure~\ref{fig:sim} (d) represents the result of a simulation with the visual extinction of 7.5~magnitude.
Our simulations suggested that carbon-chain molecules tend to decrease more slowly in the later phase.
This trend would conflict with the fact that HC$_3$N was not observed in the core C.
As shown in Figure~\ref{fig:sim}~(d), we could not find the appropriate region that explain observed column density with this parameter.

On the other hand we were able to reproduce the chemical properties in cores A and B in the appropriate evolutionary phase for dark clouds (around $1\times10^5$ years).
Our results suggest that these cores may have slightly different densities which might have lead to slightly different evolutionary phases and subsequently a little different chemical properties.
Such a slight density difference might be achievable by enhancing a very small density fluctuation through gravitational instabilities even if the L1147 filament was formed at a given time.

While the chemical properties in the core C was reproduced in a lower
density using the same visual extinction with core A and B, we could not find
the suitable region assuming lower visual extinction. This would mean that an
improvement of the model is required to better explain the chemical properties
in the core C.
In any case it was found that a combination of observational studies and chemical reaction network simulations can provide with
a powerful means in investigating chemical evolution of molecular clouds.
We would like to stress that the simulation code as well as the reaction databases are available online, and any researchers
are able to utilize this powerful tool.

Finally, although there should be a density profile in a core and the density will increase in the actual cloud evolution, our model assumed constant hydrogen density.
Thus it would be needed to develop the chemical reaction network simulation code to incorporate more realistic physical evolution
of a cloud in order to better understand the chemical evolution of a molecular cloud.
%
%

In the present study, we had to assume the excitation temperatures because we conducted single line observations.
Such observational constraint could introduce some uncertainties to our results.
Future multi-line observations would be needed to discuss the chemical differences in the early phase of cloud evolution in more detail.

\section{Conclusions}
%
%
The main results of this paper can be summarized as follows.

\begin{enumerate}
\item  It was confirmed that the L1147 cloud is a CCPR by combining the new CCS mapping result and past NH$_3$ observational results by \cite{Hirota09}.

\item  Two CCS cores and a HC$_3$N core were found in the L1147.
Three peaks are away from the Spitzer source (VeLLO source) and are  located interleaved to each other.
Such a different spatial distributions between CCS and HC$_3$N have not been known except for only one dark cloud (1172D).

\item The column densities of CCS are 3--7$\times10^{12}$~cm$^{-2}$ 
and those of HC$_3$N are 2--6$\times10^{12}$~cm$^{-2}$, respectively, which are much lower than those reported in other CCPRs.
%
%
In the three cores detected in L1147, the column densities are different within a factor of a few.
We found that the ``CCS/HC$_3$N'' ratios in the three cores are different despite that they are formed in the same filament. 

\item We used the chemical reaction network model to discuss if we can reproduce the ``CCS/HC$_3$N'' ratios derived in our observations assuming different densities from 10$^4$~cm$^{-3}$ to 10$^6$~cm$^{-3}$.
We found that our results can be reproduced for $6\times10^4$--$6\times10^5$ years; such a time range is typical for dark cloud cores and would be appropriate for CCPRs.
%
%
%
It would be likely that the difference of the ``CCS/HC$_3$N'' ratios in L1147 is explained by slightly different H$_2$ density.
L1147 would be in physically and chemically early phase of a cloud evolution, and our result suggests a possibility that chemical features in the early phase may change among cores with different physical conditions.
On the other hands,
core C is not reproduced in our simulation under realistic visual extinction. Further
development of the chemical model would be required.
%
%

\end{enumerate}

\section{Acknowledgements}
We are grateful all the staff members of Nobeyama Radio Observatory, the National Astronomical 
Observatory of Japan (NAOJ), for their support throughout our observations.
We thank to Dr. J.~M.~Kirk for consulting with our questions regarding the SCUBA map, 
to Drs. V.~Wakelam and Y.~Aikawa for providing us with the KIDA data and the chemical reaction simulation code,
and to the anonymous referee who provided us with many valuable comments..
A part of the data analysis was made at the Astronomy Data Center, NAOJ.
We utilized the Japanese Virtual Observatory (http://jvo.nao.ac.jp/) in finding the SCUBA 850~$\micron$ data.

\clearpage

\clearpage

\clearpage


\begin{deluxetable}{ccccccccc}
\tabletypesize{\scriptsize}
\tablecaption{Observed Line Characteristics towards Cores A, B, and C}
\tablewidth{0pt}
\tablehead{
\multicolumn{4}{c}{CCS ($J_N$=4$_3$-3$_4$)}  & 
\multicolumn{4}{c}{HC$_3$N ($J$=5-4)}  \\ \cline{2-5}  \cline{6-9} \\
\colhead{core} 
& \colhead{\shortstack {$T_{\rm B}$ (rms) \\ (K)} }
& \colhead{\shortstack {$V_ \mathrm{LSR}$ \\ (km~s$^{-1}$)} }  
& \colhead{\shortstack {$\Delta v$ \\ (km~s$^{-1}$)}} 
& \colhead{\shortstack {$\int T_{\rm B}$ d$v$ (3$\sigma$) \\ (K~km~s$^{-1}$)}} 
& \colhead{\shortstack {$T_{\rm B}$(rms) \\ (K)} }
& \colhead{\shortstack {$V_ \mathrm{LSR}$ \\ (km~s$^{-1}$)} }  
& \colhead{\shortstack {$\Delta v$ \\ (km~s$^{-1}$)}} 
& \colhead{\shortstack {$\int T_{\rm B}$ d$v$ (3$\sigma$) \\ (K~km~s$^{-1}$)}}
}

\startdata
A & 0.72 (0.10) & 2.6 & 0.5 & 0.38 (0.16) & 0.49 (0.09) & 2.4 & 0.5 & 0.26 (0.14)  \\
B & 0.63 (0.10) & 2.8 & 0.6 & 0.40 (0.19) & 0.95 (0.08) & 2.7 & 0.7 & 0.71 (0.18) \\
C & 0.89 (0.10) & 2.7 & 0.5 & 0.47 (0.16) & $<$0.21 (0.07) & -- & -- & $<$0.13 \\
\enddata
\label{table1}
\end{deluxetable}


\begin{deluxetable}{ccccccc}
\tabletypesize{\scriptsize}
\tablecaption{Column Densities, Fractional abundances, and CCS/HC$_3$N Ratios \break{} towards Cores A, B, and C}
\tablewidth{0pt}
\tablehead{
\colhead{core} 
& \colhead{{\shortstack{N[CCS] \\ (10$^{12}$~cm$^{-2}$)}}} 
& \colhead{{\shortstack{N[HC$_3$N] \\ (10$^{12}$~cm$^{-2}$)}}} 
& \colhead{{\shortstack{N[H$_2$] \\ (10$^{21}$~cm$^{-2}$)}}} 
& \colhead{{\shortstack{X[CCS] \\ (10$^{-10}$)}}} 
& \colhead{{\shortstack{X[HC$_3$N] \\ (10$^{-10}$)}}} 
& \colhead{N[CCS]/N[HC$_3$N]}
} 
\startdata
A & 5.8~($\pm$3.9) & 1.8~($\pm$0.9) & 6.2~($\pm$1.9) & 9.5~($\pm$7.1) & 3.0~($\pm$1.7) & 3.2~($\pm$2.6) \\
B & 3.5~($\pm$2.5) & 6.1~($\pm$4.3) & 11~($\pm$1.9) & 3.3~($\pm$2.4) & 5.7~($\pm$4.1) & 0.6~($\pm$0.6) \\
C & 7.4~($\pm$4.3) & $<$0.8 & $<$5.7 & $>$13 & - & $>$9\\
\enddata
\label{table2}
\end{deluxetable}

\begin{deluxetable}{cccc}
\tablecaption{Initial Elemental Abundances}
\tablewidth{0pt}
\tablehead{
\colhead{Element} & \colhead{Abundance} & \colhead{Element} & \colhead{Abundance}
}
\startdata
He & 9.00(-2) & Fe$^+$ & 2.00(-7) \\
N & 7.60(-5) & Na$^+$ & 2.00(-7)\\
O & 2.56(-4)  & Mg$^+$ & 2.40(-6) \\
C$^+$ & 1.20(-4)  & Cl$^+$ & 1.8(-7)\\
S$^+$ & 1.50(-5)  & P$^+$ & 1.17(-7) \\
Si$^+$ & 1.70(-6) & F$^+$ & 1.8(-8) \\
\enddata
\tablecomments{Elemental abundance used in our chemical reaction model. 
This table is referred from \cite{Wakelam08}.
}
\label{table3}
\end{deluxetable}

\end{document}